\documentclass[twocolumn]{aastex61}

\usepackage{verbatim, graphicx,  ifthen, pbox}
\usepackage{eso-pic}
\usepackage{CJKutf8}
\usepackage{xcolor}
\usepackage{hyperref}
\usepackage{breakurl}
\usepackage{amsmath}
\usepackage{graphicx}
\usepackage{verbatim}
\usepackage{booktabs}
\usepackage[T1]{fontenc}
\usepackage{ae,aecompl}

\shortauthors{Belkovski, Becker, Howe, Malsky, Batygin}
\shorttitle{HIP 41378 f's Interior Structure}
\begin{document}
\title{A multi-planet system's sole super-puff: exploring allowable physical parameters for the cold super-puff HIP 41378 f}

\author[0000-0003-4030-233X]{Michelle Belkovski}
\affiliation{College of Literature, Science, and the Arts, University of Michigan, Ann Arbor, MI 48109, USA}
\author[0000-0002-7733-4522]{Juliette Becker}
\affiliation{Division of Geological and Planetary Sciences, California Institute of Technology, Pasadena, CA 91125, USA}
\author{Alex Howe}
\affiliation{NASA Goddard Space Flight Center, 8800 Greenbelt Rd, Greenbelt, MD 20771, USA}
\author[0000-0003-0217-3880]{Isaac Malsky}
\affiliation{Department of Astronomy, University of Michigan, Ann Arbor, MI 48109, USA}
\author{Konstantin Batygin}
\affiliation{Division of Geological and Planetary Sciences, California Institute of Technology, Pasadena, CA 91125, USA}

\begin{abstract}
The census of known exoplanets exhibits a variety of physical parameters, including densities that are measured to span the range from less dense than styrofoam to more dense than iron. These densities represent a large diversity of interior structures. 
Despite this staggering diversity, recent analyses have shown that the densities of planets that orbit a common star exhibit remarkable uniformity. A fascinating exception to this is the system HIP 41378 (also known as K2-93), which contains a super-puff planet, HIP 41378 f, as well as several planets with more typical bulk densities. The range of densities in this system begs the question of what physical processes are responsible for the disparate planetary structures in this system. In this paper, we consider how the densities of the planets in the HIP 41378 system would have changed over time as the host star evolved and the planets' atmospheres were subsequently affected by the evolving insolation level. We also present a range of allowable core masses for HIP 41378 f based on the measured planet parameters, and comment on the feasibility of the proposed existence of planetary rings around HIP 41378 f as an explanation for its current low density. 
\end{abstract}
 
\section{Introduction} \label{sec:intro}
Despite the large number of exoplanets that have been physically characterized, the physics of planet formation remains shrouded in mystery. In general, the only observable piece of the processes that shape planet formation is the final product. In order to understand these processes, we must first understand the range of planet properties that can be produced by those processes. Of particular interest are the planets residing at the extremes of allowable physical properties: for example, those planets at the physical inner or outer edges of systems, those planets with the largest or smallest masses, or those planets with the highest or lowest densities. 

Understanding the planets that form at the extremes of measured densities will provide an important constraint on the processes of planet formation. 
{Planetary densities are determined using measurements of planetary radii and masses. When inferred from observations, the radius depends on more detailed physical parameters: small envelope masses can greatly inflate a planet's radius \citep{Lopez2013, Lissauer2014}; the bulk metallicity affects the atmospheric opacity for larger planets \citep{Burrows2007}; and the temperature \citep{Owen2016} affects the atmospheric dynamics. In these ways, otherwise similar planets can have substantially different densities, and a single planet can have different values of observed radius at different times.}

\begin{figure}
 \includegraphics[width=3.4in]{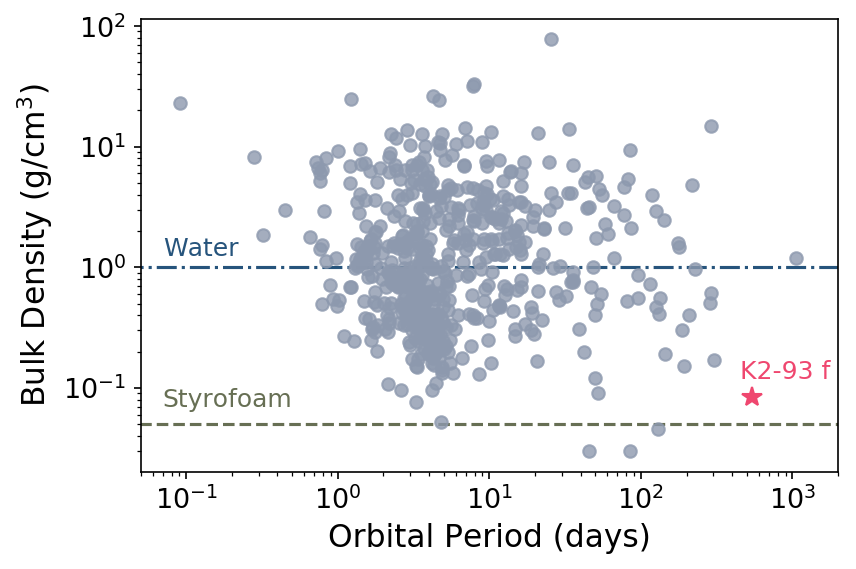}
 \caption{Bulk densities for planets where both radius and mass have been measured (the first via transit photometry, the second via either radial velocities or transit timing variations). HIP 41378 f, the main subject of analysis in this work, is marked as a pink star. Reference lines for water and Styrofoam are included. Data downloaded from NASA Exoplanet Archive on 3/1/2021.}
    \label{fig:densities}
\end{figure}

The least dense exoplanets {observed so far} appear to have bulk densities less than 0.1 g/cm$^{3}$, which can be compared with Styrofoam (0.05 g/cm$^{3}$). {Some of these planets are referred to as super-puffs \citep{Lee2016}.} In Figure \ref{fig:densities}, we plot the exoplanets for which bulk densities have been measured. 
{Of the planets known to have the lowest bulk densities, some have theoretical explanations for their densities \citep{JontofHutter2019}. For example, low-density planets with short orbital periods and the largest masses (such as WASP-107 b, \citealt{Anderson2017, Piaulet2021}; WASP-127 b, \citealt{Lam2017}; HAT-P-67 b, \citealt{Zhou2017}; HAT-S-62 b, \citealt{Hartman2019}) are characterized by their large equilibrium temperatures, suggesting that their large radii are impacted by their significant stellar insolation \citep{Guillot2002, Laughlin2011, Thorngren2018}} {or alternatively by tidal inflation \citep{Millholland2019}.}

A separate super-puff archetype is seen in Kepler-51, a system containing three super-puff planets of relatively small masses \citep{Steffen2013}. Low planetary densities can be observed when heat drives strongly outflowing atmospheres with small dust grains \citep{Wang2019}, which may be produced by a photochemical haze \citep{Gao2020, Ohno2021}. This occurs more readily for smaller, younger planets. In the case of Kepler-51, the star is young \citep{Masuda2014}, and the transmission spectra of the inner and outer planets in the Kepler-51 system appear to support the existence of a haze layer or dusty outflow \citep{LibbyRoberts2020}.

Finally, a more perplexing system architecture is found in the Kepler-79 system \citep{JontofHutter2014} and in the the HIP 41378 system \citep[also known as K2-93,][]{Vanderburg2016, Lund2019, Becker2019, Berardo2019}, where only one planet in the multi-planet system is a super-puff. Remarkably, these systems both subvert the `peas in a pod' archetype seen in the Kepler sample \citep{Weiss2018, Weiss2020}, where the ratios of adjacent planetary radii are roughly uniform within a system. This architecture, common in the exoplanet sample, is proposed to be a favorable energetic state imposed during formation \citep{Adams2019}; as the planets in the Kepler-79 and HIP 41378 systems are both under the mass limit where this architecture is expected to develop \citep{Adams2020}, the fact that both systems containing super-puffs subvert this trend may indicate alternative formation pathways as compared to the general multi-planet sample \citep{JontofHutter2019}. 

{For both Kepler-79 and HIP 41378, the old age of both these host stars, combined with the fact that the super-puff is not the innermost planet in both systems, challenge theories that work for other super-puff planets.} {Some other possible explanations for the low observed densities of super-puff planets in general include large internal heat fluxes from Ohmic dissipation \citep{Pu2017} or tidal heating \citep{Millholland2019}, planet formation that takes place further out in the disk past the ice line \citep{Chachan2021} in a cold, dust-free environment \citep{Lee2016}, or the} presence of rings in a geometry that mimics the transit of a larger planet \citep{Akinsanmi2020, Piro2020}. 
In the case of HIP 41378 f, the combination of its long orbital period \citep[$\sim$ 542 days,][]{Santerne2019} and subsequently low stellar insolation level, relatively old age \citep[2.1 $\pm$ 0.3 Gyr,][]{Lund2019}, large radius (9.2 $\pm\ 0.1\ R_{\oplus}${)}, and low mass (12 $\pm\ 3\ M_{\oplus}$, \citealt{Santerne2019}), paint a puzzle that challenges our understanding of the least dense cold planets. 

{P}revious work studying feasible equation of states for low mass planets has indicated that very low planetary densities are possible for particular combinations of core mass and envelope properties.
\citet{Batygin2013} and \citet{Chen2016} model exemplar planets and find physically consistent solutions that can allow very different radii for a planet of a given mass, depending on the core mass and other physical properties (such as metallicity). The inclusion of a wider range of variable parameters and atmospheric properties allows for the reproduction of a larger range of mass-radius combinations, even those that initially seem to have infeasibly low densities \citep{Lopez2012}. 

In this paper, we discuss where HIP 41378 f resides within the range of allowable physical planetary parameters and assess the necessity of the {various hypotheses} for HIP 41378 f's measured density. We do this by first, in Section \ref{sec:sec2}, describing feasible current states of the system that result in a planetary mass and radius consistent with the observed values. In Section \ref{sec:sec3}, we extrapolate backwards and consider how planetary mass loss due to irradiation from an evolving star would affect all planets in the HIP 41378 system. In Section \ref{sec:disc}, we discuss the degeneracies on the structure of HIP 41378 f and ways that future observations might improve the limits we provide. We then conclude in Section \ref{sec:conc} with a statement of how HIP 41378 f {contributes to} our understanding of the lowest density planets. 

\section{Current State of the System} \label{sec:sec2}

{The HIP 41378 system has a range of planetary densities, which appear to roughly decrease with orbital radius \citep{Santerne2019}.}
The inner two planets (HIP 41378 b and c, which have densities of 2.2 and 1.2 g/cm$^{3}$) have densities which appear typical for the exoplanet sample.
The next two planets in the system (HIP 41378 g and d) do not have precisely measured densities. The next planet, HIP 41378 e, has a slightly low density of 0.6 g/cm$^{3}$. 
The outermost planet in the system, HIP 41378 f, has a bulk density of 0.09 g/cm$^{3}$, among the lowest in the observed exoplanet sample. 
In this work, we do not consider HIP 41378 g, which is a low signal-to-noise ratio detection and is only detected in the RVs of \citet{Santerne2019}. As HIP 41378 g does not have a measured radius or density, we cannot place meaningful constraints on its physical structure. 
A summary of the best estimates of the orbital parameters for the remaining five planets in the HIP 41378 system can be found in Table \ref{tab:system_parameters}.

%
%
%
%
%
\subsection{Estimated Envelope Fractions for HIP 41378 b, c, d, and e}
Except for HIP 41378 f, the other planets in the HIP 41378 system have bulk densities that are easily explained by typical mass-radius relations such as \citet{Lopez2012}.
As the densities of HIP 41378 b, c, and e are fairly typical, it is straightforward to compute ranges of possible envelope fractions that match their observed masses and radii. 
HIP 41378 d has a measured radius, but only an upper limit on its mass, and so only a lower limit on the envelope fraction can be computed.

\begin{figure}
\centering
 \includegraphics[width=3.4in]{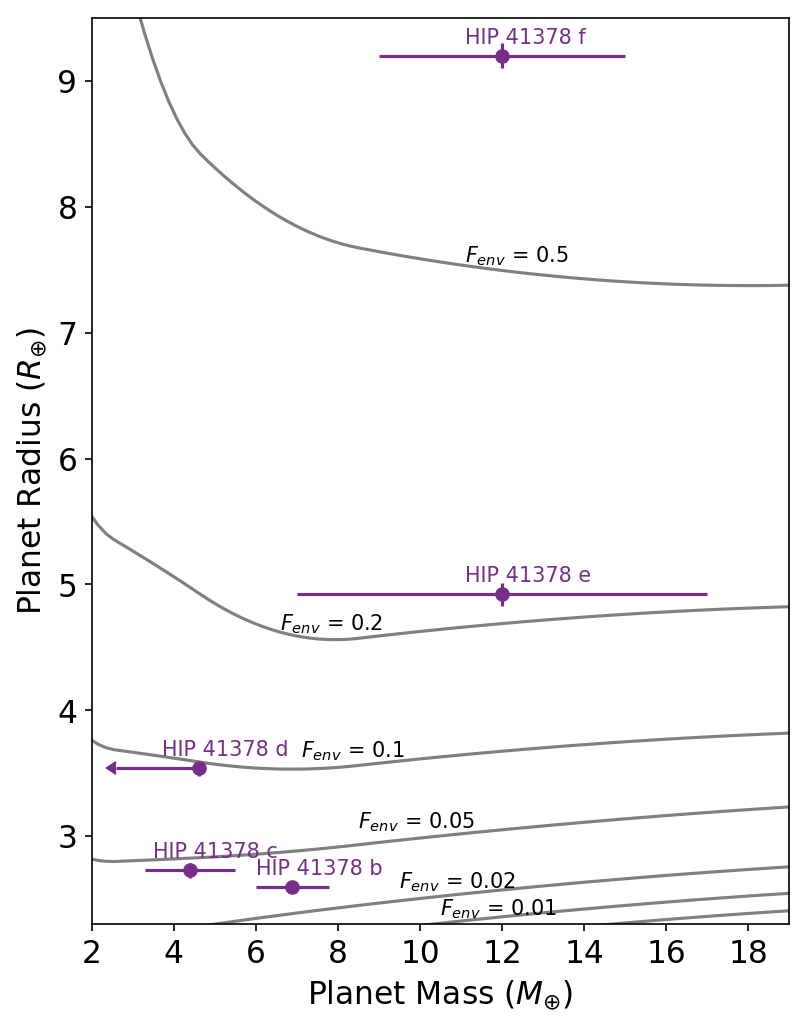}
 \caption{The measured masses and radii for the five planets with mass and radii measurements in the HIP 41378 system. Candidate HIP 41378 g is excluded because it is detected only in the radial velocity data of \citep{Santerne2019} and no radius has been measured. For HIP 41378 d, only an upper mass limit is provided, which corresponds to the 95\% credible upper limit \citep{Santerne2019}. Contour lines for a planet with solar metallicity, an age of $\sim2$ Gyr, 1 $F_{\oplus}$, and varying envelope fractions are overlaid. The models for these lines come from \citep{Lopez2012}.   }
    \label{fig:innerfour}
\end{figure}

To determine the expected envelope fraction assuming rocky cores and H/He envelopes, we {use} the models from \citet{Lopez2012} {to construct interpolated tables that allows us to model planets up to envelope fractions of 50\%. }
Using these grids, we find solutions for the current envelope fraction that match the observed mass and radius for each planet by interpolating the grids. 
The results are shown in Figure \ref{fig:innerfour} overlaid with contour lines that represent the mass-radius solution for several envelope masses at 1 $F_{\oplus}$ (which is roughly correct for the outer two planets and low for the inner two planets). The contours are produced for a solar metallicity, and an age of $\sim2$ Gyr.  The full interpolated solutions for the current-day envelope fractions based on the best estimates of the planet parameters are presented in Table \ref{tab:system_parameters}, with errors that correspond to solutions that match the observational uncertainties in mass and radius.

We note that while the mass and radius measurements reported for HIP 41378 b and c are fairly secure, the measured mass for HIP 41378 e is based on an estimate rather than a direct detection of the orbital period \citep{Santerne2019}, as the planet transited only once in the two campaigns of available K2 data \citep{Becker2019}. This value may need to be revised if an additional transit observation is made. 

\subsection{A {Hydrostatic} Solution for HIP 41378 f}
\label{sec:sec2pt2}
HIP 41378 f's density is less consistent with the range of accepted planetary densities, as HIP 41378 f has a radius of 9.2 $\pm\ 0.1\ R_{\oplus}$ \citep{Santerne2019} and a mass of 12 $\pm\ 3\ M_{\oplus}$ \citep{Santerne2019}. The combination of the mass of a mini-Neptune and the radius of a Jupiter-like planet creates a compositional puzzle.
Upon a glance at Figure \ref{fig:innerfour}, it is clear that HIP 41378 f has a much larger envelope fraction compared to the other four modeled planets. Indeed, this planet cannot be well-reproduced by the relations used above, as its measured radii is larger than any of the models from \citep{Lopez2012}. 

To determine whether it is necessary to invoke a non-standard explanation for HIP 41378 f's density, we attempted to solve for feasible parameters that create a {hydrostatic} planet while matching the current-day radius and mass observed in \citet{Santerne2019} for HIP 41378 f. To explore what these parameters might be, {most} importantly its core mass, we solved for planetary hydrostatic equilibrium using the code and method of \citet{Howe2014} and \citet{Howe2015}. 
We fix the planetary mass to be 12 $M_{\oplus}$, consistent with observations, and then {numerically} compute radii for that mass that correspond to various interior structures. 
We use an {updated} version of the code of \citet{Howe2014}\footnote{{The updated code is available at \url{https://github.com/alexrhowe/PlanetSolver}; v0.5 beta was used for this work.}}, which allows us to compute mass-radius relations for an extensive range of potential planet parameter combinations, thus offering a way to test the extreme values that a super-puff planet’s parameters typically have. 

The details of the numerical {method} are given in the appendices of \citet{Howe2014}. In short, we solve the equations of hydrostatic equilibrium using a fourth-order Runge-Kutta {method} with a combination of two equations of state: {an equation of state for one of three possible core compositions (a pure Fe core, a perovskite core, or an ice VII core)} and the \citet{Saumon1995} tabular EOS for a H$_{2}$-He envelope.
The Vinet EOS used in the core is defined as:
\begin{equation}
\begin{split}
    P = &\ 3 K_{0} (\rho / \rho_{0})^{2/3} \left( 1 - (\rho_{0} / \rho)^{1/3} \right) \times\\
    &\exp\left[1.5 (K_{0}' - 1)\left( 1 - (\rho_{0} / \rho)^{1/3} \right)\right]
   \end{split}
\end{equation}
where $P$ is the pressure, and the numerical coefficients are a reference density at zero pressure $\rho_{0}$, bulk modulus $K_{0}$, and the pressure derivative $K_{0}'$. 
{For the iron core, $\rho_{0} = 8.267$ g cm$^{-3}$, $K_{0}$ = 1.634 Mbar, and $K_{0}'$ = 5.38 Mbar \citep{Dewaele2006}; for the perovskite core, $\rho_{0} = 4.064$ g cm$^{-3}$, $K_{0}$ = 2.48 Mbar, and $K_{0}'$ = 3.91 Mbar \citep{Tsuchiya2004}; and for the ice core $\rho_{0} = 1.4876$ g cm$^{-3}$, $K_{0}$ = 1.49 Mbar, and $K_{0}'$ = 6.2 Mbar \citep{Wolanin1997}. }
For a range of planet envelope entropy values and envelope metallicities, we solve the equations of hydrostatic equilibrium with the above equations of state to determine if feasible solutions exist and what the radii of those solutions should be. 

\citet{Howe2014} found that using an Fe/rock core as opposed to an ice core does not generally significantly change the allowable envelope fraction for planets of a given mass and radius (see their Table 2 - for planets with large envelope masses, the envelope mass may change by 1 $M_{\oplus}$ dependent upon the core type). In this work, we {test three possible core compositions for HIP 41378 f under the assumption that there is} no metallicity gradient in the envelope.

\begin{figure}
 \includegraphics[width=3.5in]{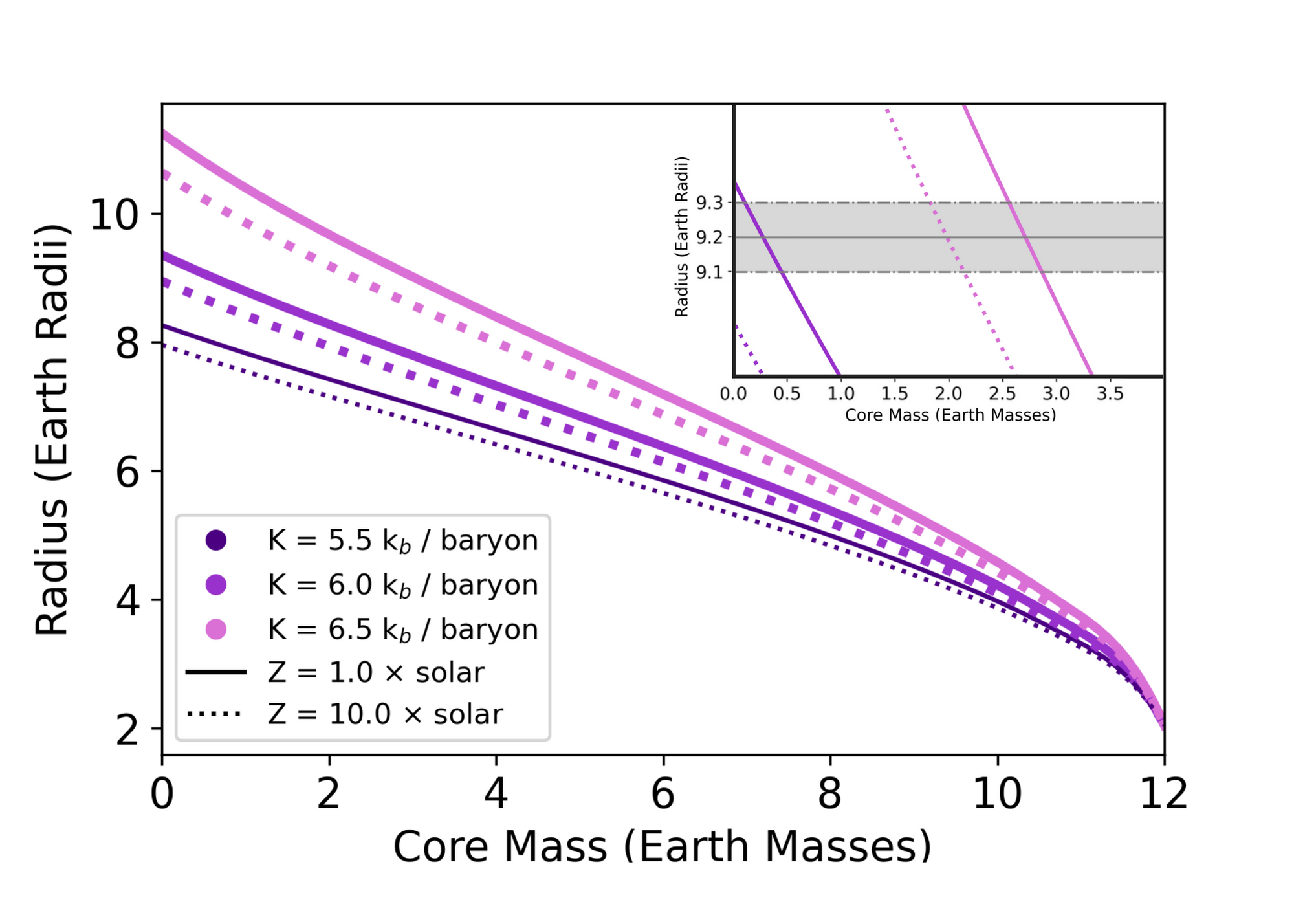}
 \caption{Mass-Radius relation for HIP 41378 f calculated with models varying in entropy and metallicity {for a planet with an iron core}. The six models plotted consist of three {entropies} (5.5, 6.0, or 6.5 {$k_{b}$/baryon}) where each entropy value was tested with a metallicity of {solar (Z=1) and 10 times solar (Z=10)}. {In the inset,} the gray shaded box represents HIP 41378 f's observed radius of 9.2 $\pm$ 0.1 R$_{\oplus}$. Three of the six models achieved this radius {by an age of 2 Gyr}, serving a better idea of what core mass range HIP 41378 f may have given the entropy and metallicity values tested in those models. From left to right: Model 1 (entropy of 6.0, metallicity of 1.0), Model 2 (entropy of 6.5, metallicity of 10.0), and Model 3 (entropy of 6.5, metallicity of 1.0).}
 \label{fig:data}
\end{figure}

To explore the potential mass-radius relations for a planet of HIP 41378 f's mass, we computed six main models {for each core type} under the above prescription, each with a varying combination of entropy (either 5.5, 6.0, or 6.5 $k_{b}$ / baryon, {the range considered in \citealt{Howe2014}}) and metallicity (either solar or 10 times solar). For each of these six models, we evaluated the interior structure of a planet with one of fourteen discrete different core mass fractions. 
For each combination of entropy, metallicity, {core composition}, and core mass, we solve for the hydrostatically stable solution and derive the planetary radius that corresponds to that solution if it exists. 
In Figure \ref{fig:data}, we show interpolated versions of the core mass - planetary radius relationship {for the case of a planet with an iron core}. All points were computed for a planet with a mass of 12 $M_{\oplus}$. 
This result shows a wide range of possible values for the radius of the planet. The primary driver of the variability in allowed radii is the core mass, with larger core masses corresponding to planets with  overall smaller radii. Secondary factors which affect the planetary radius are the entropy and metallicity of the envelope, with higher envelope entropy corresponding to puffier planets, and higher metallicity corresponding to slightly smaller radius planets.

While these relationships are intuitive, the important result is the point at which the observed radius can be reproduced and the physical consistency of those solutions with the core accretion paradigm. 
By evaluating the range of HIP 41378 f’s core mass within its given radius 9.2 $\pm$ 0.1 $R_{\oplus}$, we found the range of core masses of HIP 41378 f which reproduce the observed planetary density. 

The allowable parameters for each model {and each core type} are summarized in Table \ref{tab:cores}.
{The iron and perovskite core compositions result in very similar ranges of allowable core masses that result in planets with radii consistent with the observationally-derived value. For each model, the ice composition allows a slightly more massive core. }
{Among the three models, Model 3 ($k = 6.5 k_{b}$/baryon and Z = 1 $\times$ solar) allows for the largest possible core masses, so we consider this model moving forward. The 1$\sigma$ upper limit on core mass is 2.7 $M_{\oplus}$, 2.9 $M_{\oplus}$, and 3.1 $M_{\oplus}$ for the iron, perovskite and ice cores, respectively. Moving forwards, we take a 25\% core mass, 75\% envelope mass to be representative of these solutions; however, it is possible that the core mass is much less massive (as would be the case in Models 1 and 2). }

\begin{table*}
  \begin{center}
    \caption{Orbital periods, planetary radii, and measured planetary eccentricities come from \citet{Santerne2019}. Inferred properties (initial and final envelope mass and fraction) come from the results of this work. $^{*}$: value taken for a planet with a current day envelope entropy of 6.5 $k_{b}$/baryon consistent with the $3\sigma$ lower limit of the planet radius {for the Fe cores, which is also roughly consistent with the $1\sigma$ upper limits for the perovskite and ice cores.} Based on results of Section \ref{sec:sec2pt2}, {depending on the core composition,} this value could range between 74\% and 99\%+.  }
    \label{tab:system_parameters}
    \begin{tabular}{p{6.5cm}p{1.9cm}p{1.9cm}p{1.9cm}p{1.9cm}p{1.9cm}} 
{Parameter} & {HIP 41378 b}	&	{HIP 41378 c}	&	{HIP 41378 d} & {HIP 41378 e} & {HIP 41378 f} \\
\hline
{Orbital Period (days)} & {15.57208 $\pm$ 2 $\times$ $10^{-5}$} & {31.70603 $\pm$ 6 $\times$  $10^{-5}$} & {278.3618 $\pm$ 5 $\times$  $10^{-4}$} & {369 $\pm$ 10} & {542.07975 $\pm$ 1.4 $\times$  $10^{-4}$} \\
{Semi-major Axis (AU)} & 0.1283 & 0.2061 & 0.88 & 1.06 & 1.37 \\
{Eccentricity $e$} & 0.07 & 0.04 & 0.06 & 0.14 & 0.004 \\
{Radius ($R_{\oplus}$)} & 2.595 & 2.727 & 3.54 & 4.92 & 9.2 \\
{Mass ($M_{\oplus}$)} & 6.89 & 4.4 & $<$4.6 & 12.0 & 12.0 \\
{Inferred Initial Envelope Mass ($M_{\oplus}$)} & 0.42 & 0.36 & 0.47 & 2.75 & 9 \\
{Inferred Final Envelope Mass ($M_{\oplus}$)} & {0.21} & 0.19 & 0.45 & 2.74 & 9$^{*}$ \\
{Inferred Final Envelope Fraction ($m_{env}$ / $m_{p}$)} & 3\% $\pm$ 1\% & 4\% $\pm$ 1\% & $>$10\% $\pm$ 1\% & 22\% $\pm$ 5\% & $>$75\% 
\label{tab:params}
    \end{tabular}
  \end{center}
\end{table*}

\section{System State at Formation} \label{sec:sec3}
In Section \ref{sec:sec2}, we found plausible planetary structures for all five planets with measured masses and radii that are consistent with the current-day observed values. Due to the fact that the inner two planets have the highest densities in the system and the importance of photo-evaporation in sculpting the radius distribution of close-in planets \citep{SanzForcada2011, Owen2017, Uzsoy2021}, we next consider whether the system once consisted of planets with higher envelope mass fractions that diminished over time due to the radiation from their host star, HIP 41378, as it evolved. In this scenario, the innermost planets would be most strongly affected by the stellar XUV radiation and experience the largest fractional loss of the envelopes, potentially resulting in the density gradient that we see today. 

\subsection{General Analysis of Mass Loss for HIP 41378 b, c, d, e}

\begin{figure}
\centering
 \includegraphics[width=3.4in]{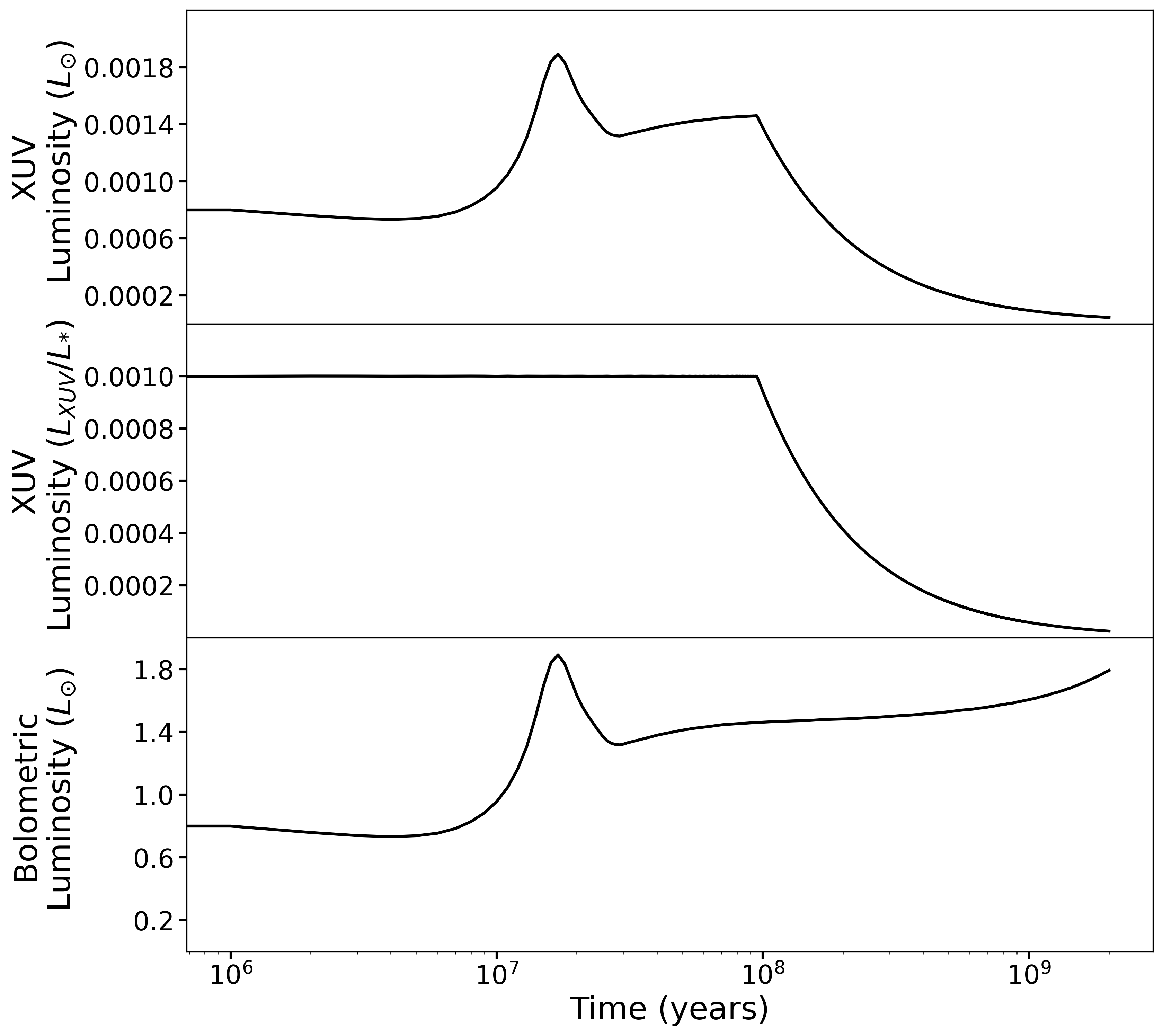}
 \caption{The stellar luminosity of HIP 41378 in (top panel) the XUV, (middle panel) the XUV luminosity as a fraction of the total luminosity, and (bottom panel) the full bolometric luminosity. The XUV luminosity drives photoevaporative mass loss from the planetary envelopes. }
    \label{fig:stellar}
\end{figure}

We used the program \texttt{VPLanet}\footnote{{Version 2.2, available at \url{https://github.com/VirtualPlanetaryLaboratory/vplanet}}} \citep{Barnes2019} to model the evolution of the five planets with measured masses and radii. \texttt{VPLanet} allows us to solve how the envelope mass of each planet in the system changes as the star evolves. The mass of a planet will decrease over time due to XUV radiation from the host star. 
{XUV radiation from the host star will heat a planetary atmosphere via the ionization of atmospheric H/He, and the heat will drive atmospheric outflows \citep{VidalMadjar2003, Lammer2003, Owen2012, MurrayClay2009}. }
{This process} will occur at a higher rate with a higher incident XUV flux. In Figure \ref{fig:stellar}, we show the evolution of the luminosities as the star ages, as computed by \texttt{VPLanet} for a 1.16 $M_{\odot}$ star \citep[which is the mass of the host star HIP 41378;][]{Santerne2019} and based on the stellar models of {\citet{Baraffe2015}}. The XUV luminosity, which drives mass loss, is strongest when the star is young. 

We compute the mass loss due to XUV radiation over two billion years using the \texttt{atmesc} and \texttt{stellar} modules along with the parameters set in Table \ref{tab:params}. 
We use variable time-stepping with an accuracy coefficient of 0.01, and integrate between $5 \times 10^{6}$ and $2 \times 10^{9}$ years using the \citet{Baraffe2015} stellar model and assume a saturated XUV luminosity fraction of $10^{-3}$ which persists for the first 0.1 Gyr of the star's lifetime \citep{Lammer2014}. The planetary semi-major axis and mass values are set to the best-fit values from \citet{Santerne2019} and summarized in Table \ref{tab:system_parameters}. We use \texttt{VPLanet}'s automatic atmospheric escape prescriptions, which switches between radiation/recombination-limited \citep{Luger2015} and energy-limited escape according to the incident XUV flux. 
{In contrast to energy-limited escape, which occurs at lower incident FUV fluxes, some of the inputted energy is lost via radiative cooling in recombination-limited escape \citep{MurrayClay2009}. }
The \texttt{VPLanet} analysis works in terms of the planetary mass, with the mass-radius models of \citet{Lopez2012} being subsequently used to convert the mass into a radius. 
As discussed in Section \ref{sec:sec2}, the \citet{Lopez2012} mass-radius models describe well the structures of all planets except for HIP 41378 f. We note that the radius cannot accurately be computed for HIP 41378 f using this model, an issue we will return to in Section \ref{sec:sec3part2}. {We also note two additional caveats: (1) We assume that the stellar X-ray and EUV fluxes will evolve on the same timescale, but this may not always be true \citep{King2021}; (2) The efficiency factor for photoevaporative mass loss will depend on the specific level of XUV radiation as well as the planet gravity \citep{Caldiroli2022}, so further refinement of the planet masses or stellar luminosity could alter our results. }

\begin{figure}
\centering
 \includegraphics[width=3.4in]{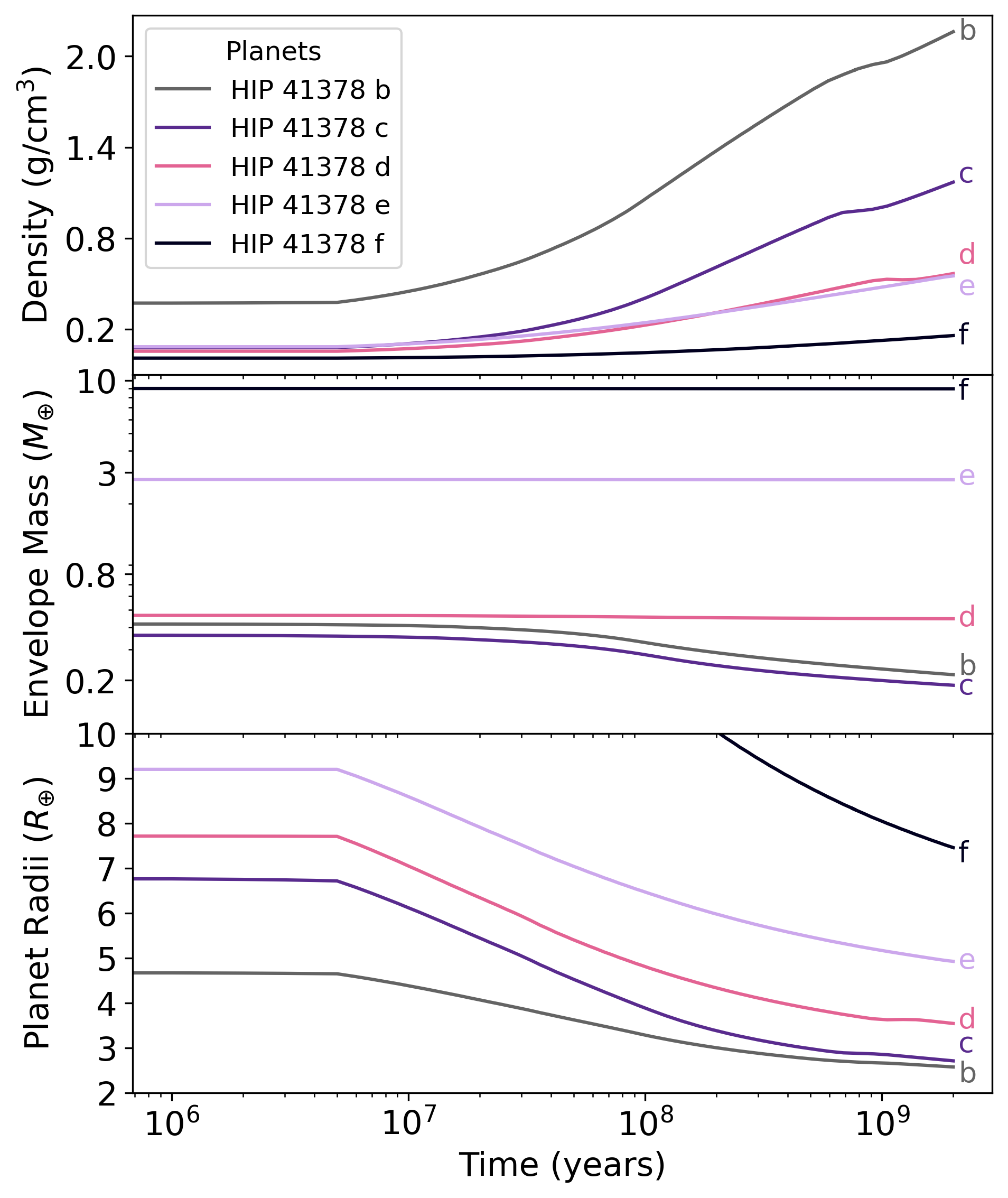}
 \caption{Subplots of the HIP 41378 planets' density, envelope mass, and radius values over 2 Gyr. All planets in the system have increasing densities and decreasing radii over time. However, planets HIP 41378 b and HIP 41378 c have envelope masses that  decrease {by a factor of two over the simulated period}.}
    \label{fig:rmdsubplot}
\end{figure}

While we have the final current-day values for the mass, radius, and expected envelope fraction for each planet, we want to test what feasible envelope fractions could have been at formation.
To test this and determine which HIP 41378 planets may have had significantly higher envelope mass fractions at formation, we perform a parameter space survey where we vary the estimated initial envelope masses for each planet in the system and then compute the mass loss over 2 Gyr, attempting to match the current value. Once we found which initial envelope mass fractions led to the integration output matching current day observed values, we could then estimate by how much each planet’s envelope mass fraction likely diminished over two billion years, the estimated current age of the system \citep{Lund2019}. {Once the parameter space survey yielded a result consistent with the currently measured system parameters, we computed a full system evolution with \texttt{VPLanet}. The resultant evolutions of the planetary radii, envelope masses, and densities are shown in Figure \ref{fig:rmdsubplot}.}

\begin{figure}
\centering
 \includegraphics[width=3.4in]{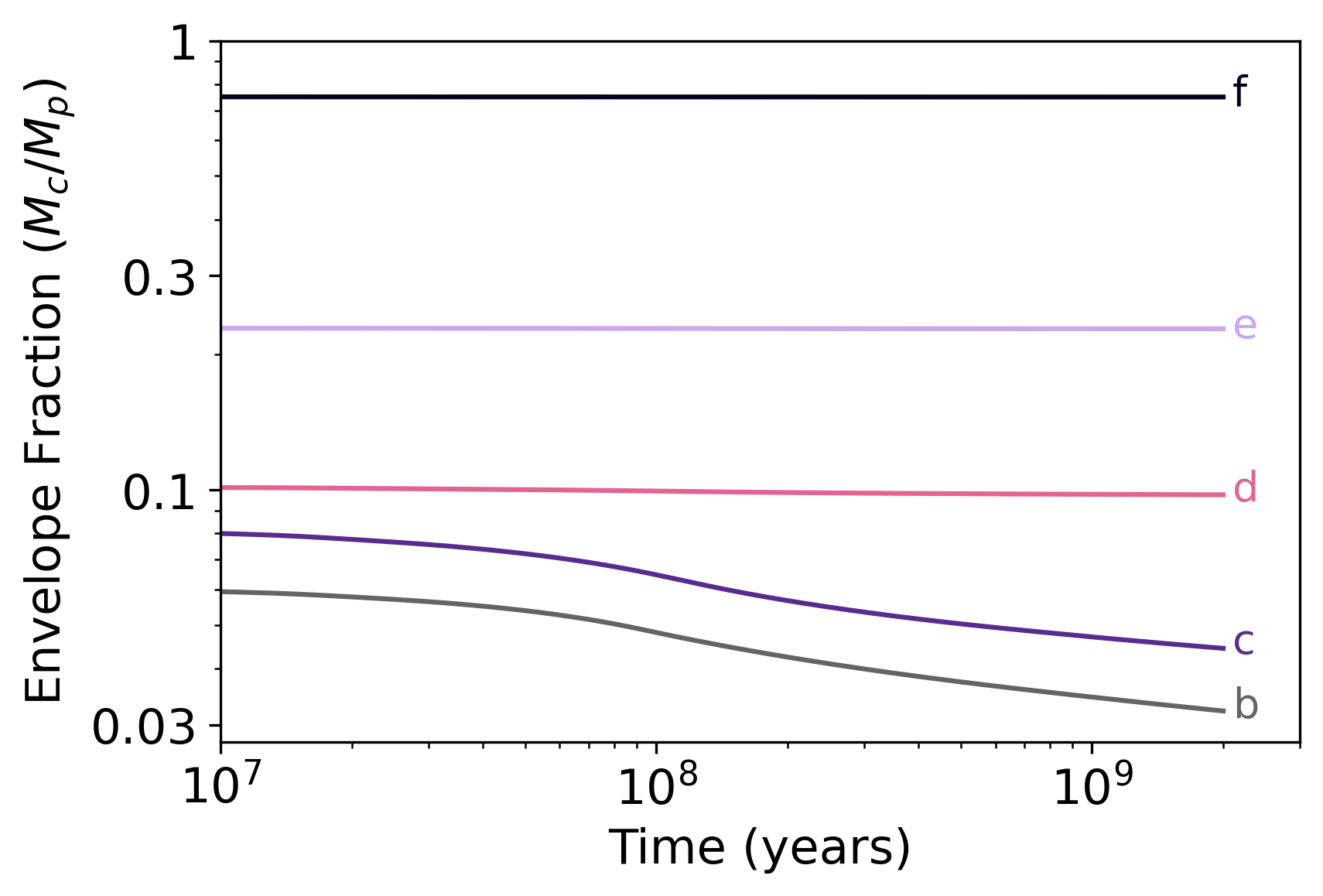}
 \caption{The envelope mass fractions of each planet in HIP 41378 over 2 Gyr, computed with \texttt{VPLanet}. While planets HIP 41378 d, HIP 41378 e, and HIP 41378 f's envelope mass fractions are essentially constant over time, HIP 41378 b and HIP 41378 c's values decrease by roughly a factor of 2 {by an age of 2 Gyr}.}
    \label{fig:envelopefraction}
\end{figure}

We {also show a zoomed in view of the planetary envelope mass fraction} over time in Figure \ref{fig:envelopefraction}.
We observe that the two inner planets HIP 41378 b and HIP 41378 c had initial envelope masses that substantially decreased due to the XUV luminosity from the system’s host star.
HIP 41378 d, e, and f lose very minimal amounts of mass from their envelopes as the star evolves, due to their larger orbital separations from the host star. 
Our \texttt{VPLanet} analysis supports our initial hypothesis that the two inner planets in the HIP 41378 system likely had low densities in the past, and lost some of their envelopes due to the XUV radiation from the host star. 
However, the amount of XUV radiation on the atmospheres of HIP 41378 b and c is not expected to have been enough to drive the evolution from an initial super-puff density to the present-day value. Instead, it is likely that HIP 41378 b and c had at formation {envelope fractions a factor of two larger than their present day values}. Similarly, the middle planets in the system (HIP 41378 d and HIP 41378 e) likely did not experience much XUV-driven photoevaporation, and their current-day observed envelope fractions are likely very similar to the values at formation.

\begin{table*}
\caption{Feasible values for HIP 41378 f's core mass given varying combinations of entropy and metallicity that result in planets with the observed present physical radii within $1 \sigma$ of the observationally determined value \citep[$9.2 \pm 0.1 R_{\oplus}$;][]{Santerne2019}. }
\centering 
\begin{tabular}{c c c c c c c} 
\hline\hline 
Model\ & Entropy \ & Metallicity\ & Allowable Core Mass & Allowable Core Mass & Allowable Core Mass \vspace{-0.4em} \\
 &  {($K_{b}$/baryon)}& {($Z_{solar}$)} & (Iron) ($M_{\oplus}$, $1\sigma$)  & (Perovskite) ($M_{\oplus}$, $1\sigma$)  & (Ice) ($M_{\oplus}$, $1\sigma$)\\
 [0.5ex] 
\hline 
1 & 6.0 & 1.0 & [0.1, 0.4] & [0.1, 0.4]& [0.1, 0.5]\\ 
2 & 6.5 & 10.0 & [1.7, 2.0]& [1.8, 2.1]& [2.0, 2.3]\\
3 & 6.5 & 1.0 & [2.4, 2.7]& [2.6, 2.9] & [2.7, 3.1]\\[1ex] 
\hline\hline 
\label{tab:cores}
\end{tabular}
\label{table:nonlin} 
\end{table*}

\subsection{Detailed Analysis of HIP 41378 f's Evolution}
\label{sec:sec3part2}
{To further study HIP 41378 f's non-standard interior structure, we perform additional explorations of its time evolution} using coupled interior structure and atmospheric modeling of planets with Earth composition rocky cores and primordial hydrogen-helium envelopes \citep{Rogers2011}. {We} performed these calculations using the Modules for Experiments in Stellar Astrophysics (\texttt{MESA}) code \citep[v12778;][]{Paxton2011, Paxton2013, Paxton2015, Paxton2018, Paxton2019} and with the additions outlined in \cite{Malsky_2020}. The framework from \cite{Malsky_2020} specifically allows for the modeling of sub-Neptune mass exoplanets over a range of envelope mass fractions, planet core masses, and mass loss parameterizations. By matching the observed properties of HIP 41378 f to these 1D structure models, we can characterize the interior structure of the planet and how it evolves over the course of billions of years.

We parameterized the 1D evolution model in MESA as follows. {Based on the results of the previous section, which showed that the mass of HIP 41378 f did not significantly change over the system age, we choose the initial mass and envelope fraction to be equal to the values found above. As such, w}e assume an initial planet mass of 12.0 M$_\oplus$, an initial envelope mass fraction of 0.75, an orbital separation of 1.37 au, and initial abundances of $X=0.74$, $Y=0.24$, $Z=0.02$. We set the outer atmospheric boundary layer within MESA at $\tau$=2/3, and implement a grey Eddington T($\tau$) relation with the \verb|atm_T_tau_relation| option. This is the outermost zone for which MESA solves the set of coupled structure  and composition equations. We then extrapolate the atmosphere up to 1.0 mbar assuming an isothermal temperature profile and a constant atmospheric mean molecular mass. The pressure level of 1.0 mbar corresponds to the radii observed by transit surveys \cite{2009ApJ...702.1413M}. All other model parameters are identical to the full model description in \cite{Malsky_2020}, except for the removal of the `hot start' evolution step. This step was required in \cite{Malsky_2020} in order to standardize the starting entropy of highly irradiated super-Earths, and is not required for HIP 41378 f's evolution.

Our 1D models of HIP 41378 f are consistent with a bulk density of 0.087, and match observational results.  
By 2.1 Gyr, the model had a transit radius of 9.1 R$_\oplus$, consistent to $1\sigma$ with the measured value from \citet{Santerne2019}, and as expected from the previous analysis, the planet did not exhibit significant mass loss over its lifetime. The {approximate lower bound} 75\% envelope fraction derived from the {hydrostatic} solutions does reproduce the observed planet radius through the MESA modeling.

{Another concern for the atmospheric retention of HIP 41378 f is the possibility that the planet would have lost its atmosphere early due to Parker winds \citep{Parker1958}. For some planets, the Parker wind causes rapid boil-off of their atmospheres on kyr timescales \citep{Owen2016c}. For super-puff planets, Parker winds can potentially prevent the retention of a planetary atmosphere. }

{To test whether Parker winds could have caused HIP 41378 f to have lost its atmosphere in a problematically short timescale, we compute the mass loss rate using Equation 5 of \citet{Chachan2020}:}
\begin{equation}
    dM/dt = 4 \pi r_{s}^{2} c_{s} \rho_{p} \exp{(3/2 - 2 r_{s} / R_{p})},
\end{equation}
{where $r_s = G m_{p} / 2c_{s}^{2}$ is the sonic radius, $m_{p}$ the planetary mass, $c_s = \sqrt{k_{b}T / \mu}$ the isothermal sound speed, $\mu \approx 2.2$ amu the assumed atmospheric mean molecular weight, $\rho_p$ the atmospheric density at the radius of the planet, and $R_{p}$ the planetary radius. For HIP 41378 f, the equilibrium temperature is $T \approx 300$ K \citep{Santerne2019}, and we compute the speed of sound to be $c_s = 1$ km/sec, sonic radius to be $r_{s} = 0.016$ au and $\rho_p = 10^{-3}$ g/cm$^{3}$. Using these values, we find that the Parker wind mass loss rate is $\sim 10^{-7}$ g/s, corresponding to an atmospheric mass loss timescale longer than the Hubble time.}

{HIP 41378 f's large mass (12 $M_{\oplus}$) causes the Parker wind mass loss rate to be small enough that this planet would not have been expected to lose a substantial amount of its atmosphere via early boil-off. However, it is important to note that while the Parker wind mass loss is not important for HIP 41378 f's evolution, the same may not be true for other similar planets; if HIP 41378 f's mass had been 25\% of its current value, for example, its mass loss rate would been $10^{18}$ g/s and the atmosphere would have been lost on kyr timescales. In this way, HIP 41378 f's relatively large mass allowed it to retain its atmosphere \citep[similarly to other larger mass, larger gas-to-core ratio planets as in][]{Vissapragada2020}.   }

\section{Discussion} \label{sec:disc}
In this work, we have presented self-consistent solutions for HIP 41378 f's interior structure (including core mass and envelope entropy) both today and after formation (2 Gyr ago). We have also derived estimates of the current-day envelope fractions for HIP 41378 b, c, d, and e, and used \texttt{VPLanet} to simulate their evolution  {while accounting for the evolving XUV flux}, and thus estimated the initial envelope fractions. We find that while HIP 41378 b and c (with periods of roughly 15 and 30 days respectively) likely had significantly (by a factor of $\sim2$) higher envelope fractions at formation, none of the planets except HIP 41378 f formed with anomalously low densities. 

Planets with high H/He envelope mass-fractions may experience mass loss due to {large XUV fluxes} \citep{Lammer2003}. This is most relevant for planets residing close to their host stars \citep{Owen2016}, {in which case substantial initial envelope fractions may prevent the destruction of a low-density planet's planetary atmosphere \citep{Hallatt2021}, particularly for short-period planets}. 
{Short-period planets will also be affected by mass loss due to Parker winds early in their lifetimes; for some super-puff planets, this could lead the entire atmosphere to be lost on $\sim$ kyr timescales \citep{Chachan2020}. For HIP 41378 f, {the predicted mass loss rate is predicted to be very small, meaning that mass loss due to isothermal Parker winds will be negligible}. While this rate could be an underestimate if HIP 41378 f has significant internal heat flux, that is unlikely for a planet with such a large orbital radius \citep{Pu2017}.  }
In cases like HIP 41378 f, where the planet's orbital separation is large enough that mass loss will be minimal, the primary factor driving the change in physical radius from formation to today is the change in the envelope's entropy as the planet gravitationally contracts. At formation, cold planets with masses {around} 10 $M_{\oplus}$ can have {entropies} ranging between 9 - 9.5 $k_b$/baryon \citep{Mordasini2017}. 
Our investigations with PlanetSolver and the MESA simulations show that at 2 Gyr, HIP 41378 f can have {a radius} and density consistent with the observed values if the planet has an envelope entropy of 6.5 $k_b$/baryon.

Our results show that a high envelope fraction (around 75\%) and envelope entropy {of} 6.5 $k_{b}$/baryon are also consistent with HIP 41378 f's measured planetary mass and radius. A transient enhancement in planetary envelope entropy (which increases the effective radius of a planet) could be recent collisions from small rocky planetesimals or even from other planets \citep{Ketchum2011, Anderson2012}. However, the necessary mass-flux of solid materials required to enhance the planetary entropy sufficiently at current day (2 Gyr) would be several orders of magnitude larger than the current mass of the planet, and so this explanation is not feasible. 
{Similarly, we can assess the feasibility of some of the alternative hypotheses for super-puff planetary densities. Due to HIP 41378 f's relatively large (1.3 au) orbital distance, theories that depend on stellar insolation such as tidal inflation \citep{Millholland2019} or Ohmic dissipation \citep{Pu2017} can be dismissed for this particular planet. }

{Transmission spectroscopy can further differentiate between the remaining feasible theories, one of which is an outflowing atmosphere with a large population of small grains \citep{Wang2019} or photochemical hazes \citep{Gao2020} which may produce dust grains \citep{Ohno2021}. This explanation, although more likely to affect younger planets, will be testable with transmission spectroscopy of HIP 41378 f in the future, as such dusty atmospheres are expected to have flat transmission spectra \citep{Kawashima2019}. 
{However, we compute very low expected mass loss rates for HIP 41378 f via both photoevaporation and Parker winds, which makes it unlikely that dust would exist at altitudes sufficient to obscure transmission spectra. }
In the case of Kepler-79, {water features are not detected in the transmission spectrum} \citep{Chachan2020}, supporting the haze hypothesis. In contrast, if transmission spectra show evidence for water features in HIP 41378 f's atmosphere, that {could} be evidence in favor of the hypothesis of \citet{Lee2016} {(see also \citealt{Lee2018})}, in which super-puffs form  dust-free in a cold formation environment far from the host star. Indeed, the largest core mass is possible with an ice core, and if HIP 41378 f did have an ice core, this would place additional constraints on its formation process. An ice core requires very particular formation conditions: the planet likely would have formed past the ice line in the disk, then migrated inwards - an origin possibly consistent with the scenario outlined in \citet{Lee2016}. However, recent {HST} results from \citet{Alam2022} suggest  no evidence for water features in HIP 41378 f's transmission spectrum.}

A core mass fraction of 25\% or less for a planet of HIP 41378 f's mass also appears immediately inconsistent with the paradigm of giant planet formation. Planets this small are unlikely to form via disk instability {\citep{Boss1997, Dodson-Robinson2009, Boley2009}}, so the most likely pathway to a planet like HIP 41378 f is via core accretion. In the core accretion model, cores form via accretion of planetesimals. Following the formation of the core, the planet will accrete nearby gas until the mass of the core and the mass of the envelope are roughly equal, at which point runaway accretion can occur and the envelope may become far more massive. Barring the absence of a sufficient reservoir of gas, the outcome of runaway accretion is generally a Jupiter-mass planet {\citep{Pollack1996}}. 
All of our {hydrostatic} solutions require envelope fractions for HIP 41378 f of 75\% or greater, {and for a planet with this core mass, the gas fraction will be set by how much the core can accrete during the disk lifetime \citep{Lee2019}. To create a planet with a mass of only 12 $M_{\oplus}$} would subsequently require the termination of accretion slightly after the start of runaway accretion (before HIP 41378 f would have had the chance to grow to a Jupiter mass). This issue could be solved with particularly serendipitous (and low probability) timing for the disk dissipation, {a solution which has also been invoked for the paucity of close-in Jupiter-mass planets \citep{Lee2014}}. However, evidence suggests that the lifetime of the protoplanetary disk is not generally the primary driver of the end of planetary mass accretion \citep{Adams2021} 

There is a second, more fatal flaw with the core accretion picture for HIP 41378 f: the allowable core mass we compute is smaller than the limit required to accrete a significant ($>50\%$) envelope fraction, no matter where in the disk the planet formed. The critical core mass for runaway accretion is estimated to be upwards of 5-10 $M_{\oplus}$ \citep{Papaloizou1999, Rafikov2006, Coleman2017}, but we compute a maximum core mass of 3 $M_{\oplus}$ with the value being only 0.4 $M_{\oplus}$ under more standard assumptions about the planetary entropy. Since this is well below the threshold required to trigger runaway accretion, there is no easy way to explain the 75\%+ envelope fraction in the core accretion paradigm. 

In this paper, we attempted to find a feasible structure for HIP 41378 f which does not require invoking the existence of planetary rings. Although {hydrostatic} solutions do exist, these solutions are not consistent with the paradigm of core accretion due to the combination of core mass and envelope fraction. 
Because of this, there is no easy explanation for how HIP 41378 f could have formed with its measured parameters.
Our analysis provides support for the hypothesis of \citet{Piro2020} and \citet{Akinsanmi2020} that HIP 41378 f's anomalously low density could be due to the presence of planetary rings. As discussed in \citet{Piro2020}, HIP 41378 f is the only known super-puff planet for which the rings hypothesis could be assessed using ground-based telescopes. 
{Simultaneously with the review process of this manuscript, \citet{Alam2022} announced HST results showing a flat transmission spectrum for HIP 41378 f. \cite{Ohno2022} also demonstrated that flat transmission spectra are expected for planets with ringed geometries, supporting the potential for rings in the case of HIP 41378 f. }
It is important to continue to monitor HIP 41378 f for TTVs and refine its transit ephemeris \citep[as in][]{Bryant2021} {so that future JWST observations can be made}. 

\section{Conclusion} \label{sec:conc}
This work considers both the current envelope fractions and feasible past evolutions of envelope fractions for the planets in the HIP 41378 system, with a particular focus on the super-puff HIP 41378 f. We find estimated current-day envelope fractions by solving the equations of hydrostatic equilibrium, then use \texttt{VPLanet} to deduce their envelope fractions at formation. We find that the inner two planets (HIP 41378 b and c) did not have anomalously low densities at formation, but that HIP 41378 f did. 

Although HIP 41378 f's structure can be explained by a combination of larger envelope entropy values and high envelope fractions, there remain problems with the theoretical interpretation of those values. Namely, as discussed in this paper, consistent solutions for HIP 41378 f's interior structure require current-day envelope fractions upwards of 75\%, which provides a tension with the core accretion paradigm. Because of this, {ways in which the planetary radius may have overestimated by the transit data should be considered, including the theory advanced in \citet{Piro2020} and \citet{Akinsanmi2020} that the anomalously large radius of HIP 41378 f could be due to planetary rings. This particular theory appears consistent with current data \citep{Alam2022, Ohno2022}, and will be directly testable with additional observations with JWST. }

\medskip
\textbf{Acknowledgements.} Michelle Belkovski thanks the University of Michigan Undergraduate Research Opportunity and Research Scholars programs for their support of this project. J.C.B.~has been supported by the Heising-Simons \textit{51 Pegasi b} postdoctoral fellowship.
We thank Rodrigo Luger and Rory Barnes for useful conversations and assistance with VPLanet. We also thank the anonymous referee for their invaluable insight into our paper which led to substantial improvements. This research has made use of the NASA Exoplanet Archive, which is operated by the California Institute of Technology, under contract with the National Aeronautics and Space Administration under the Exoplanet Exploration Program.

\facility {Exoplanet Archive}
\software{pandas \citep{mckinney-proc-scipy-2010}, IPython \citep{PER-GRA:2007}, matplotlib \citep{Hunter:2007}, scipy \citep{scipy}, numpy \citep{oliphant-2006-guide}, Jupyter \citep{Kluyver:2016aa}, \texttt{VPLanet} \citep{Barnes2019}, PlanetSolver \citep{Howe2015} }

\bibliographystyle{mnras}
\bibliography{refs}

\end{document}